# Solution to the Counterfeit Coin Problem and its Generalization

**J. Dominguez-Montes**
Departamento de Físca, Novavision, Comunidad de Canarias, 68 - 28230 Las Rozas (Madrid)
www.dominguez-montes.com
jdm@nova3d.com

***Abstract*:**

This work deals with a classic problem: "Given a set of coins among which there is a counterfeit coin of a different weight, find this counterfeit coin using ordinary balance scales, with the minimum number of weighings possible, and indicate whether it weighs less or more than the rest". The method proposed here not only calculates the minimum number of weighings necessary, but also indicates how to perform these weighings, it is easily mechanizeable and valid for any number of coins. Instructions are also given as to how to generalize the procedure to include cases where there is more than one counterfeit coin.

## *1. Introduction*

Information Theory was created in 1948 by Claude Shannon [(1)]. This theory has notably enriched the field of research into mathematics, economics, biology, psychology, semantics, etc. As an example, this theory recently contributed to quantum mechanics [(2)]. This work seeks to provide the definitive solution to the problem of finding a counterfeit coin in a set n, characterized by having a different weight, using ordinary scales, with the smallest number of weighings. This problem has been widely discussed by many authors, although none has offered a systematic solution. Yaglon and Yaglon[(3)], among others, widely discuss this problem in the light of Information Theory. However, they do not offer an easily mechanizeable or generalizeable method.

The author [(4)] published a first work regarding the solution of this problem in the Spanish magazine *Qüestiió* in an article which details only the case with 12 coins and 3 weighings. The current version has been extended to include an example of generalization for the case of 4 weighings and 39 coins. Parts of sections 1, 2 and 3 are maintained as in the first publication and contain an introduction to Information Theory which many readers may wish to skip, and go directly to section 5, where the detailed analysis of the problem's solution begins.

## *2. Measuring the amount of information*

The first and most important contribution to Information Theory was made by Shannon and Wiener, who evidenced the statistical nature of communications. Accordingly radio, television, teletypes and any other information source choose "randomly", but with a given probability, the sequence of messages of a specific vocabulary.

Thus, once a message has been received, we do not know which will be the next message, since its choice will be random, but we do know the probability of transmitting each message directly. If the choice were deterministic, the message would not contain any information, since its choice would be known beforehand.

Consequently, it seems reasonable to suppose that our search of a measurement of the amount of information should be restricted to finding a statistical parameter associated with the schema of probabilities.

Shannon suggested that the random variable $-log\ p(x_k)$ was the measurement of information associated with the occurrence of an event $x_k$. He showed that the average value of this function is a good indication of the average value of the uncertainty which derives from performing a random experiment. This average amount of information is known as Entropy. Entropy can be interpreted as being the average amount of information per symbol emitted by a source.

When we choose the base of logarithms equal 2 the choice of one of two equiprobable events gives us an amount of information unit commonly known as the Bit. When the base of the logarithms is 10 the information unit corresponds to the choice of one among ten equiprobable cases and is known the Hartley.

### *3. Hamming codes capable of correcting a single error*

This coding procedure shows how it is possible to make a reality of the transmission of information in the presence of noise fulfilling the minimum redundancy predicted by Information Theory.

Let us imagine a binary message of $n$ positions. The amount of information $I_e$ necessary to discover which of the $n$ elements of value 0,1 is erroneous would be:

$$Ie = -log\ \frac{1}{n+1} = log\ (n+1)$$

In other words, the choice of 1 among $n+1$ equiprobable elements. A number $n$ of possible different positions for the error, and an additional position to indicate that the message is correct.

A number $k$ of binary elements 0,1 of redundancy or parity provide an amount of information $I_k$:

$$I_k = -k\ log\ 1/2 = log\ 2^k$$

Evidently, the amount of information provided by the digits of parity must be sufficient to know where the error is located, wherefrom: $I_k \geq Ie$ or:

$$2^k \geq n+1$$

Let us imagine that we want to transmit a binary message with $n$ = 7 positions. The number $k$ of bits among them which must be used as parity bits must at least three, since:

$2^3 = 7+1$

The simplest way to implement a coding procedure that fulfills the above requirement is to choose as parity bits those which occupy positions 1, 2 and 4 and the following parity equations (the sign + indicate the sum module 2):

| | |
|---|---|
| $x_1 + x_3 + x_5 + x_7 = par$ | $s_1$ |
| $x_2 + x_3 + x_6 + x_7 = par$ | $s_2$ |
| $x_4 + x_5 + x_6 + x_7 = par$ | $s_3$ |

With the above parity equations it is easy to decide which he erroneous bit is. For example, if there is a fault in:

| | |
|---|---|
| $s_1$ | the error is bit $x_1$ |
| $s_2$ | the error is bit $x_2$ |
| $s_3$ | the error is bit $x_4$ |

If there are faults in:

| | |
|---|---|
| $s_1$ and $s_2$ | the error is bit $x_3$ |
| $s_1$ and $s_3$ | the error is bit $x_5$ |
| $s_2$ and $s_3$ | the error is bit $x_6$ |
| $s_1$ $s_2$ and $s_3$ | the error is bit $x_7$ |

The above parity equations are shown in a more compact way via the parity matrix *H*.

| | $x_1$ | $x_2$ | $x_3$ | $x_4$ | $x_5$ | $x_6$ | $x_7$ | |
|---|---|---|---|---|---|---|---|---|
| | 1 | 0 | 1 | 0 | 1 | 0 | 1 | $s_1$ |
| $H =$ | 0 | 1 | 1 | 0 | 0 | 1 | 1 | $s_2$ |
| | 0 | 0 | 0 | 1 | 1 | 1 | 1 | $s_3$ |

Each column of the matrix is the binary representation of the order number of each column and each row a parity equation.

The choice of the bits occupying positions 1, 2 and 4 as parity bits is owing to the fact that, since they appear only once in the parity equations, they can easily be deduced as a function of the info bits.

For example, after codifying message 1 1 0 0 with the corresponding parity bits, this will become message $x_1$ $x_2$ 1 $x_4$ 1 0 0 and bits $x_1$, $x_2$ and $x_4$ will be obtained from the equations:

| | | | | | |
|---|---|---|---|---|---|
| $x_1 = x_3 + x_5 + x_7$ | = | 1+1+0 | = | 0 | $s_1$ |
| $x_2 = x_3 + x_6 + x_7$ | = | 1+0+0 | = | 1 | $s_2$ |
| $x_4 = x_5 + x_6 + x_7$ | = | 1+0+0 | = | 0 | $s_3$ |

Giving the complete message: 0 1 1 1 1 0 0.

If, for example, the fifth bit were erroneous, the message received would be 0 1 1 1 0 0 0 and it would be easy to see that the binary vector $s_3 \; s_2 \; s_1$ would be 1, 0, 1 which in the matrix corresponds to the order position number 5:

$$\begin{aligned} x_1 + x_3 + x_5 + x_7 &= 0 + 1 + 0 + 0 = 1 \quad s_1 \\ x_2 + x_3 + x_6 + x_7 &= 1 + 1 + 0 + 0 = 0 \quad s_2 \\ x_4 + x_5 + x_6 + x_7 &= 1 + 0 + 0 + 0 = 1 \quad s_3 \end{aligned}$$

Thus, we would be operating with an error in any other position.

With another additional parity bit in position eight, we could build codes able to detect and correct an error in a message of length 15 and so on.

There are no codes with this correction capability and less redundancy than the Hamming codes, which are therefore known as perfect codes(5).

### *4. The counterfeit coin problem*

With $n$ coins all the same weight except for one which could weigh more or less, determine the minimum number of weighings $x$ which must be performed in balance scales to identify whether this coin exists and whether it is heavier or lighter than the rest.

The scales have three positions: left-tilted, right-tilted or balanced. Consequently, the amount of information $I_p$ obtained in each weighing will be:

$$I_p = \log 3$$

If we perform $x$ weighings, the total information obtained will be:

$$xI_p = \log 3^x$$

Furthermore, we must determine which coin among $n$ is different we therefore need an amount of information:

$$I_1 = \log n$$

and since the coin must also weigh more or less the additional amount of information:

$$I_2 = \log 2$$

The total amount of information necessary to determine which coin is counterfeit and whether it is heavier or lighter than the rest will be:

$$I_1 + I_2 = \log 2n$$

If we also accept the additional case that all coins may be good, we need additional information equivalent to the increase deriving from changing from selecting one among $2n$ to one among $2n+1$. The total necessary amount of information $I_T$ will be:

$$I_T = log\ (2n+1)$$

Of $x$ weighings, we must obtain sufficient information to decide whether or not there is a counterfeit coin and whether it is heavier or lighter than the rest, thus must be complied with:

$$xI_p \geq I_T$$

hence,

$$3^x \geq 2n+1$$

If $n$ number of coins is 12 the value of $x$ must be at least 3 since:

$$3^3 \geq 24+1$$

Although using three weighings we obtain an amount of information: $3^3 = 27$ that is sufficient to identify one counterfeit coin among 13; (2•13+1=27), the use of weighing scales to obtain the information means that each weighing or linear equation must contain an even number of coins and therefore that the total number of coins has to be a multiple of 3, this restriction reduces the number of coins with 3 weighings from 13 to 12, with 4 weighings from 40 to 39, etc., as shown in more detail in section 7.

If a counterfeit coin were known to exist, and its weight were known, the amount of information from $x$ weighings would be sufficient for a number $n$ of coins given by $n=3^x$.

### *5. Weighing procedure to identify counterfeit coin*

Information Theory and Coding [6] show us how to code messages in the best possible way in order to be able to detect and correct errors in information transmission or storage.

There is an evident analogy between codes able to detect and correct **"1"** error in a message of **"n"** digits and the procedure to find **"1"** counterfeit coin and discover whether it weighs more or less in a set of **"n"** coins [4].

No other code can match the Hamming codes capacity of detecting and correcting errors and none has less redundancy, which is why their application to the counterfeit coin problem will provide us with the best procedure for weighing the coins.

In order to implement a Hamming code, a redundancy or test matrix is constructed. This matrix has as many rows as redundant equations and as many columns as digits in the message.

The way to perform the weighings can be expressed by a table of numbers in which each row represents one weighing and each column a different coin. The Hamming test matrix can therefore be interpreted as the table of numbers which expresses the way in which the weighings must be performed.

Since the coin may weigh the same, more, or less than the rest of the coins, we will have 3 possible values which we will call 0, 1 and 2. Just as the balance scales can adopt three different positions, balanced, tilted leftward, and tilted rightward, we will again have three possible values which we will call 0, 1 and 2.

Accordingly, we must build a table of numbers using only values 0, 1 and 2, i.e. in base 3, following the same procedure as for building a test matrix for written messages in base 3.

The procedure for building this Hamming test matrix or, what amounts to the same, the table of numbers to indicate the manner in which to perform the weighings, is as follows:

The table of numbers must have as many rows as weighings, for example, if we assume three weighings it must have three rows. The numbers appearing in the table must be written with the digits 0, 1 and 2, i.e. in base 3. These three-digit numbers will be $3^3 = 27$ following numbers: 000, 001, 002, 010, 011, 012, 020, 021, 022, 100, 101, 102, 110, 111, 112, 120, 121, 122, 200, 201, 202, 210, 211, 212, 220, 221, 222.

Of these, 000 must be eliminated, since it would indicate that the coin does not appear in any of the weighings, and we are left with a total of $3^3$-1 different numbers.

If each of these must represent a different coin, it is necessary to prevent the sum of any two vectors or columns in base three from being nil, as in columns (0, 1, 2) and (0, 2, 1). This difficulty may be overcome if all “the first elements” other than zero of each of the columns of the test matrix *H* are 1. Thus, no pair of columns are linearly interdependent and this matrix is valid to build codes capable of detecting and correcting an error.

Of the total number of columns different from zero that can be written with *x* rows in base 3; $3^x - 1$, it will be necessary to choose only one fraction 1/(3-1) whose first non-nil element is equal to 1. The maximum column number obtained in this way is:

$$\frac{3^x - 1}{2}$$

And this number must be equal to or higher than that of the coins:

$$\frac{3^x - 1}{2} \geq n$$

In the particular case, if a counterfeit coin is known to exist and its weight is known, compliance with $3^x \geq n$ will be sufficient.

In the general case, for *n=12* coins, the number of weighings or linear equations or rows in matrix *H* must be equal to or higher than 3. The test matrix for *x=3* is:

| | 1 | 2 | 3 | 4 | 5 | 6 | 7 | 8 | 9 | 10 | 11 | 12 | 13 | |
|---|---|---|---|---|---|---|---|---|---|---|---|---|---|---|
| | 0 | 0 | 0 | 0 | 1 | 1 | 1 | 1 | 1 | 1 | 1 | 1 | 1 | $s_1$ |
| *H* = | 0 | 1 | 1 | 1 | 0 | 0 | 0 | 1 | 1 | 1 | 2 | 2 | 2 | $s_2$ |
| | 1 | 0 | 1 | 2 | 0 | 1 | 2 | 0 | 1 | 2 | 0 | 1 | 2 | $s_3$ |

The correction of errors will be performed in the same way as in the binary case. Test equations $s_1$ $s_2$ and $s_3$ are verified and, if they are all complied with, there is no error and, therefore, no counterfeit coin. If any one is not complied with, then the first non-nil value of $s_1$ or $s_2$ or $s_3$, will indicate error *b*, and vector ($s_1/b$, $s_2/b$, $s_3/b$) will indicate its position.

We applied the above test matrix to the problem of the weighings. In the first place, we will ignore column 13, since there are only 12 coins.

The linear equation $s_1$ will indicate which coins must intervene in the first weighing. This can be the comparison of coins 5, 6, 7, and 8 on an "arbitrary" side of the scales, for example, the left, which we call 1, and coins 9, 10, 11, and 12 on the right; which we call 2, we will interpret zero value as these coins not figuring in the weighing. Thus:

First weighing $s_1$

**5, 6, 7, 8 <> 9, 10, 11, 12**

The result of this comparison will give a value for $s_1$ which could be 0, 1 or 2, according to whether the scale balances, tilts leftward or tilts rightward.

For the second weighing, we will also use matrix *H*. The elements other than zero in the second row of matrix H will indicate the coins which must appear in the second weighing. It should also be considered that the coins that appear in the linear equation $s_2$ with a 2 must change side of the scales; this happens with coins 11 and 12, which must be on the left, since in the first weighing they were on the right, and those whose value does not change such as 8, 9 and 10 must remain on the same side of the scales. Of the rest of the coins, those with the highest value, 3 and 4, must be assigned to whichever side of the scales has been called 2. Thus:

Second weighing $s_2$

**2, 8, 11, 12 <> 3, 4, 9, 10**

and the same method must be followed for the third weighing, i.e. 6 and 12 must remain on the same side as in weighing 1, and 3 and 9 must remain on the same side as in weighing 2, and 7 and 10 must change with respect to weighing 1 and 4 must change with respect to weighing 2. Thus:

Third weighing $s_3$

**1, 4, 6, 10 <> 3, 7, 9, 12**

The values obtained in the first weighing $s_1$, second weighing $s_2$ and third weighing $s_3$, will be the components of vector $s_1, s_2, s_3$ which will identify the counterfeit coin if it exists and, if so, will pinpoint whether it weighs more or less than the others. The steps to follow for its identification are as follows:

A) If $s_1$ is different from 0 note said value as divisor $b$ and obtain the vector quotient $s_1/b, s_2/b, s_3/b$ and go to step B. Otherwise, continue in M.

   M. If $s_2$ is other than 0 note said value as divisor $b$ and obtain the vector quotient $s_1/b, s_2/b, s_3/b$ and go to step B. Otherwise, continue in N.

   N. If $s_3$ is other than 0 note said value as divisor $b$ and obtain the vector quotient $s_1/b, s_2/b, s_3/b$ and go to step B. Otherwise, all the coins are good, so stop.

B) The result of step A gives us a vector quotient $s'_1\ s'_2\ s'_3 = (s_1/b, s_2/b, s_3/b)$ and a divisor b whose value can either be 1 or 2. with the vector quotient, the counterfeit coin's location in the *H* matrix will be identified and we will proceed to step C.

C) Value 1 or 2 of the divisor $b$ obtained in step A serves to determine whether the coin weighs more or less. Since value 2 was assigned arbitrarily for movement of the scales to the right and the coins were also distributed arbitrarily between the left and right sides of the scales in the first coin, we must interpret the divisor 2. In the above case of arbitrary assignment and distribution of this value 2 indicates that the counterfeit coin is heavier if it turns out to be 3, 4, 9, 10, 11, 12 because the first time these appeared in a weighing they were in the right hand side of the scales and lighter if the counterfeit coin is 1, 2, 5, 6, 7, 8. In contrast, if the divisor is 1, this will indicate less weight for coins 3, 4, 9, 10, 11 and 12, and more weight for coins 1, 2, 5, 6, 7 and 8.

Example 1:

Supposing the result of the three weighings is as follows:

First weighing
    Moving to the right    $s_1 = 2$
Second weighing
    Moving to the right    $s_2 = 2$
Third weighing
    Moving to the right    $s_3 = 2$

In step A, we obtain the vector quotient 1, 1, 1 and the divisor $b = 2$.

In step B we can see via matrix *H* that vector 1, 1, 1, indicates position 9 and therefore that this is the counterfeit coin.

Step C indicates that divisor 2 obtained in step A indicates for coin 9 that it is heavier.

Example 2:

Supposing that the result of the weighing were:

First weighing
In balance $s_1 = 0$
Second weighing
In balance $s_2 = 0$
Third weighing
Movement to the right $s_3 = 2$

In step A, we obtain the vector quotient 0, 0, 1 and the divisor $b = 2$.
In step B we can see via matrix H that vector 0, 0, 1 corresponds to the coin in position 1.
In step C we see that divisor 2 obtained in step A means for the first coin 1 that it is lighter.

Example 3:

Supposing that the result of the weighing had been:

First weighing
To the left $s_1 = 1$
Second weighing
To the left $s_2 = 1$
Third weighing
In balance $s_3 = 0$

In step A, we obtain the vector quotient (1, 1, 0) and the divisor $b$=1.
In step B we can see via matrix *H* that vector (1, 1, 0) corresponds to position 8. That is to say, the counterfeit coin is number 8.
In step C we can see that divisor 1 means that coin number 8 is heavier.

### *6. Practical procedure for the case of 12 coin*

The test matrix must comply with the condition that the sum module 3 of any two vectors or columns is not nil. Any matrix meeting this condition can be valid.

For example, in the aforementioned cases, all "the first elements" other than zero in each column of the test matrix have been assigned the value 1 in order to meet this condition. A test matrix whose first elements other than zero were 2 would also work.

It might sometimes be preferable to build a more complex test matrix in exchange for the weighings being easier. Here, there is an additional condition that the test matrix must have the same number of 0s, 1s and 2s in each row in order to be able to directly assign the coins to a particular side of the scales.

In order to obtain a test matrix with these characteristics, we will use the above test matrix in which no pair of vectors or columns added module 3 is nil. We have seen that this matrix is as follows:

| 1 | 2 | 3 | 4 | 5 | 6 | 7 | 8 | 9 | 10 | 11 | 12 | 13 |
|---|---|---|---|---|---|---|---|---|---|---|---|---|
| 0 | 0 | 0 | 0 | 1 | 1 | 1 | 1 | 1 | 1 | 1 | 1 | 1 |
| 0 | 1 | 1 | 1 | 0 | 0 | 0 | 1 | 1 | 1 | 2 | 2 | 2 |
| 1 | 0 | 1 | 2 | 0 | 1 | 2 | 0 | 1 | 2 | 0 | 1 | 2 |

For each row of this matrix to have the same number of 0s, 1s and 2s, the following property will be used: "if any vector of the matrix is multiplied by 2 module 3, the sum of any pair of vectors or columns continues to be non-nil".

The first row of the matrix contains four 0s, eight 1s and no 2s. In order for it to have four 1s and four 2s, we will multiply the four vectors and columns on the right side, i.e. those corresponding to coins 9, 10, 11, 12 by 2 module 3 (this operation transforms 1s into 2s, 2s into 1s and maintains the 0 values) the matrix thereby having the same number of 0s, 1s and 2s in the first row:

| 1 | 2 | 3 | 4 | 5 | 6 | 7 | 8 | 9 | 10 | 11 | 12 |
|---|---|---|---|---|---|---|---|---|---|---|---|
| **0** | **0** | **0** | **0** | **1** | **1** | **1** | **1** | **2** | **2** | **2** | **2** |
| 0 | 1 | 1 | 1 | 0 | 0 | 0 | 1 | 2 | 2 | 1 | 1 |
| 1 | 0 | 1 | 2 | 0 | 1 | 2 | 0 | 2 | 1 | 0 | 2 |

The second row now has four 0s, six 1s and two 2s. In order for it to contain four 2s, we will multiply by 2 module 3 the two vectors with the highest values (furthest to the right) which contain 1 in row 2 and 0 in row 1. These vectors will be columns 3 and 4. The matrix will therefore be as follows:

| 1 | 2 | 3 | 4 | 5 | 6 | 7 | 8 | 9 | 10 | 11 | 12 |
|---|---|---|---|---|---|---|---|---|---|---|---|
| **0** | **0** | **0** | **0** | **1** | **1** | **1** | **1** | **2** | **2** | **2** | **2** |
| **0** | **1** | **2** | **2** | **0** | **0** | **0** | **1** | **2** | **2** | **1** | **1** |
| **1** | **0** | **2** | **1** | **0** | **1** | **2** | **0** | **2** | **1** | **0** | **2** |

The matrix thereby has the same number (four) of 0s, 1s and 2s in each row. In addition to indicating which coins must appear in each weighing, this matrix identifies with the value 1 those coins which should appear, *for example*, on the left side of the scales and with a 2 those which must appear on the right and with 0 those which do not intervene in the weighing. The following weighings can therefore be deduced:

$s_1$ **5, 6, 7, 8 <> 9, 10, 11, 12**

$s_2$ **2, 8, 11, 12 <> 3, 4, 9, 10**
$s_3$ **1, 4, 6, 10 <> 3, 7, 9, 12**

The result of the weighings $s_1$, $s_2$, $s_3$ obtained by assigning 1 to the left-hand tilt, 0 for balance and 2 for the right-hand tilt will serve to identify in the matrix which is the counterfeit coin and in this case that it is heavier.

If $s_1$, $s_2$, $s_3$, cannot be found in the matrix, the vector $s_1/2$, $s_1/2$, $s_1/2$, will be obtained (this operation is equivalent to replacing value 1 with 2 and 2 with 1 and 0 remaining unchanged) the coin will be identified in the matrix and in this case it is lighter.

Example 1:

Supposing the result of the three weighings is as follows:

First weighing:
Moving to the right $s_1=2$
Second weighing:
Moving to the right $s_2=2$
Third weighing:
Moving to the right $s_3=2$

The counterfeit coin is 9 and it is heavier.

Example 2:

First weighing
In balance $s_1=0$
Second weighing
In balance $s_2=0$
Third weighing
To the right $s_3=2$

The counterfeit coin is [$s_1=0/2=0$, $s_2=0/2=0$, $s_3=2/2=1$] 1 and lighter.

## *7. Generalization*

The generalization of the procedure for any number $n$ of coins is immediate. It is sufficient to build a test matrix whose first non-nil element is equal to 1 in each column. Each row corresponds to a linear equation or weighing, and the number $x$ of these must fulfill the inequality:

$$3^x - 1 \geq 2n$$

where $n$ will be the number of columns or coins.

We have already demonstrated that it is always possible to build such a matrix.

There is an additional restriction on applying the Hamming test matrix to the problems of coins and scales deriving from the fact that the number of coins which must be involved in each weighing must be even and therefore that the total number of coins must be a multiple of 3. If we choose the equal sign in the above inequation, the value of *n* will indicate the maximum number of columns among which the counterfeit coin can be found with *x* weighings:

$$n = \frac{3^x - 1}{2}$$

However, in order for it to be feasible to perform the weighings, this number must be a multiple of 3, which does not occur in the case *n* above, but does for the immediately inferior *n'*:

$$n' = n - 1 = \frac{3^x - 3}{2} = 3\frac{\left(3^{x-1} - 1\right)}{2} = 3p$$

Since *p* is always an integer number for any integer value of *x* higher than 1.

This means that for *x* = 3 the maximum number of coins is 12. Similarly, for *x* = 4 weighings, the maximum number of coins is 39, for x = 5 it will be 120, for *x* = 6 it will be 363, for *x* = 7 it will be 1.092, etc.

### *8. Example of generalization: four weighings and thirty-nine coins*

The maximum number n' of coins among which a counterfeit coin can be detected with *x* = 4 weighings is obtained from the expression:

$$n' = \frac{3^x - 3}{2} = \frac{3^4 - 3}{2} = 39$$

The test matrix must have as many rows as number of weighings, and in this case it will have 4 rows, and as many columns as coins, so in this case it will have 39 columns. The columns must be in base 3 and their first non-nil element must be 1. The test matrix H will therefore be:

Coins

| | | 1 | 2 | 3 | 4 | 5 | 6 | 7 | 8 | 9 | 10 | 11 | 12 | 13 | 14 | 15 | 16 | 17 | 18 | 19 | 20 |
|---|---|---|---|---|---|---|---|---|---|---|---|---|---|---|---|---|---|---|---|---|---|
| | $s_1$ | 0 | 0 | 0 | 0 | 0 | 0 | 0 | 0 | 0 | 0 | 0 | 0 | 0 | 1 | 1 | 1 | 1 | 1 | 1 | 1 |
| | $s_2$ | 0 | 0 | 0 | 0 | 1 | 1 | 1 | 1 | 1 | 1 | 1 | 1 | 1 | 0 | 0 | 0 | 0 | 0 | 0 | 0 |
| Weighings | $s_3$ | 0 | 1 | 1 | 1 | 0 | 0 | 0 | 1 | 1 | 1 | 2 | 2 | 2 | 0 | 0 | 0 | 1 | 1 | 1 | 2 |
| | $s_4$ | 1 | 0 | 1 | 2 | 0 | 1 | 2 | 0 | 1 | 2 | 0 | 1 | 2 | 0 | 1 | 2 | 0 | 1 | 2 | 0 |

| | | 21 | 22 | 23 | 24 | 25 | 26 | 27 | 28 | 29 | 30 | 31 | 32 | 33 | 34 | 35 | 36 | 37 | 38 | 39 | 40 |
|---|---|---|---|---|---|---|---|---|---|---|---|---|---|---|---|---|---|---|---|---|---|
| | $s_1$ | 1 | 1 | 1 | 1 | 1 | 1 | 1 | 1 | 1 | 1 | 1 | 1 | 1 | 1 | 1 | 1 | 1 | 1 | 1 | 1 |
| | $s_2$ | 0 | 0 | 1 | 1 | 1 | 1 | 1 | 1 | 1 | 1 | 1 | 2 | 2 | 2 | 2 | 2 | 2 | 2 | 2 | 2 |
| Weighings | $s_3$ | 2 | 2 | 0 | 0 | 0 | 1 | 1 | 1 | 2 | 2 | 2 | 0 | 0 | 0 | 1 | 1 | 1 | 2 | 2 | 2 |
| | $s_4$ | 1 | 2 | 0 | 1 | 2 | 0 | 1 | 2 | 0 | 1 | 2 | 0 | 1 | 2 | 0 | 1 | 2 | 0 | 1 | 2 |

### *9. Intuitive procedure for the case of 39 coins*

In the first place, we shall proceed as before, i.e. via an intuitive procedure.

Equation $s_1$ indicates the coins which must be involved in the first weighing and these will be those coins that in the first row of matrix $H$ have a value other than 0. The coins will be assigned such that the highest values are assigned to the side of the scales denominated by 2. This weighing could be the comparison between the following groups of coins:

**[$s_1$] 14, 15, 16, 17, 18, 19, 20, 21, 22, 23, 24, 25, 26 < > 27, 28, 29, 30, 31, 32, 33, 34, 35, 36, 37, 38, 39**

Equation $s_2$ indicates the coins which must be involved in the second weighing and these will be those coins what in the second row in matrix $H$ have a value other than 0. The coins which in the vector column appear with a 2 must change sides on the scales and those which appear with a 1 having been involved in previous weighings with the same number 1 must remain on the same side. Thus, coins 32,33,34,35,36,37,38,and 39 change sides with respect to weighing 1 and 23,24,25,26,27,28,29,30, and 31 remain on the same side The coins will be assigned such that the highest values are assigned to the side of the scales denominated by 2.

The second weighing would therefore be as follows:

**[$s_2$] 5, 13, 23, 24, 25, 26, 32, 33, 34, 35, 36, 37, 38, 39 < > 6, 7, 8, 9, 10, 11, 12, 13, 27, 28, 29, 30, 31**

Equation $s_3$ or third row of the matrix $H$, indicates the coins which must appear in the third weighing. Similarly, the coins which in the vector column change value from 1 to 2 or from 2 to 1 must change sides on the scales and those which appear with a 1 having been involved in previous weighings with the same number 1 must remain on the same side. Thus, coins 11, 12, 13, 29, 30, 31, 35, 36, and 37 change sides with respect to weighing 2 and 20, 21, 22 change sides with respect to weighing 1. Similarly, 17, 18, 19 remain on the same side as in weighing 1 and 8, 9, 10, 26, 27, 28, 38 and 39 remain on the same side as in weighing 2. The rest of coins may be assigned in such a way that the highest values are in the side of the scales known as 2..

The third weighing will therefore be:

**[$s_3$] 2, 11, 12, 13, 17, 18, 19, 26, 29, 30, 31, 38, 39 < > 3, 4, 8, 9, 10, 20, 21, 22, 27, 28, 35, 36, 37**

Equation $s_4$ or fourth row of the matrix $H$ indicates the coins which must appear in the fourth weighing. Similarly, the coins which in the vector column change value from 1 to 2 or from 2 to 1 must change sides on the scales and those which maintain the same value must remain on the same side. Thus, coin 16 will change sides with respect to weighing 1, coins 7, 10, 25 and 33 change sides with respect to weighing 2, coins 4, 12, 19, 21, 28, 30, 37 and 39 change sides with respect to weighing 3. Similarly, coin 15 will remain on the same side as in weighing 1, coins 6, 9, 24 and 34 will remain on the same side as in weighing 2 and coins 3, 13, 18, 22, 27, 31 and 36 will remain on the same side as in weighing 3. Consequently, weighing four will be:

**[$s_4$] 1, 4, 7, 10, 13, 15, 18, 21, 24, 28, 31, 34, 37 < > 3, 6, 9, 12, 16, 19, 22, 25, 27, 30, 33, 36, 39**

Since value 2 was assigned arbitrarily for moving the scales to the right and the coins were also distributed arbitrarily between the left and right sides of the scales in the first weighing, value 2 will indicate that the coin is heavier and value 1 that it is lighter if the counterfeit coin turns out

to be 4, 5, 6, 7, 8, 9, 10, 11, 12, 27, 28, 29, 30 31, 32, 33, 34, 35, 36, 37, 38, 39, since the first time that these coins appeared on the scales they were on the right and similarly value 2 will indicate that the coin is lighter and value 1 that it is heavier if the counterfeit coin turns out to be 1, 2, 3, 13, 14, 15, 16, 17, 18, 19, 20, 21, 22, 23, 24, 25, 26, since the first time these coins appeared on the scales they were on the left side.

Example 1:

Supposing that the result of the four weighings is:

$s_1 = 1, \quad s_2 = 2, \quad s_3 = 2, \quad s_4 = 1$

The vector quotient is 1, 2, 2, 1 and the divisor 1.

In the test matrix, the vector 1, 2, 2, 1 corresponds to coin 39, for which divisor 1 means that it is lighter.

Example 2:

Supposing that the result of the four weighings is:

$s_1 = 2, \quad s_2 = 2, s_3 = 1, s_4 = 1$

The vector quotient is 1, 1, 2, 2, and the divisor 2.

In the test matrix, vector 1, 1, 2, 2 corresponds to coin 31, for which divisor 2 means that it is heavier.

Example 3:

Supposing that the result of the weighings is:

$s_1 = 1, \quad s_2 = 0, \quad s_3 = 0, \quad s_4 = 1$

The vector quotient is 1, 0, 0, 1 and the divisor 1. In the test matrix, this vector corresponds to coin 15, for which divisor 1 means that it is heavier.

### *10. Practical procedure for the case of 39 coins*

The test matrix must comply with the condition that the sum module 3 of any two vectors or columns is not nil. Any matrix meeting this condition can be valid. For example, in the aforementioned cases, all “the first elements” other than zero in each column of the test matrix have been assigned the value 1 in order to meet this condition. A test matrix whose first elements other than zero were 2 would also work.

It might sometimes be preferable to build a more complex test matrix in exchange for the weighings being easier. Here, there is an additional condition that the test matrix must have the same number of 0s, 1s and 2s in each row in order to be able to directly assign the coins to a particular side of the scales.

In order to obtain a test matrix with these characteristics, we will use the above test matrix in which no pair of vectors or columns added module 3 is nil. We have seen that this matrix is as follows:

Coins

| | | 1 | 2 | 3 | 4 | 5 | 6 | 7 | 8 | 9 | 10 | 11 | 12 | 13 | 14 | 15 | 16 | 17 | 18 | 19 | 20 | 21 | 22 | 23 | 24 | 25 | 26 | 27 | 28 | 29 | 30 | 31 | 32 | 33 | 34 | 35 | 36 | 37 | 38 | 39 | 40 |
|---|---|---|---|---|---|---|---|---|---|---|---|---|---|---|---|---|---|---|---|---|---|---|---|---|---|---|---|---|---|---|---|---|---|---|---|---|---|---|---|---|---|
| | $s_1$ | 0 | 0 | 0 | 0 | 0 | 0 | 0 | 0 | 0 | 0 | 0 | 0 | 0 | 1 | 1 | 1 | 1 | 1 | 1 | 1 | 1 | 1 | 1 | 1 | 1 | 1 | 1 | 1 | 1 | 1 | 1 | 1 | 1 | 1 | 1 | 1 | 1 | 1 | 1 | 1 |
| | $s_2$ | 0 | 0 | 0 | 0 | 1 | 1 | 1 | 1 | 1 | 1 | 1 | 1 | 1 | 0 | 0 | 0 | 0 | 0 | 0 | 0 | 0 | 0 | 1 | 1 | 1 | 1 | 1 | 1 | 1 | 1 | 1 | 2 | 2 | 2 | 2 | 2 | 2 | 2 | 2 | 2 |
| Weighings | $s_3$ | 0 | 1 | 1 | 1 | 0 | 0 | 0 | 1 | 1 | 1 | 2 | 2 | 2 | 0 | 0 | 0 | 1 | 1 | 1 | 2 | 2 | 2 | 0 | 0 | 0 | 1 | 1 | 1 | 2 | 2 | 2 | 0 | 0 | 0 | 1 | 1 | 1 | 2 | 2 | 2 |
| | $s_4$ | 1 | 0 | 1 | 2 | 0 | 1 | 2 | 0 | 1 | 2 | 0 | 1 | 2 | 0 | 1 | 2 | 0 | 1 | 2 | 0 | 1 | 2 | 0 | 1 | 2 | 0 | 1 | 2 | 0 | 1 | 2 | 0 | 1 | 2 | 0 | 1 | 2 | 0 | 1 | 2 |

For each row of this matrix to have the same number of 0s, 1s and 2s, the following property will be used: "if any vector of the matrix is multiplied by 2 module 3, the sum of any pair of vectors or columns continues to be non-nil".

The first row of the matrix contains thirteen 0s, twenty six 1s and no 2s. In order for it to have thirteen 1s, we will multiply by 2 module 3 the thirteen vectors with the highest values, i.e. those furthest to the right, in this case 27, 28, 29, 30, 31, 32, 33, 34, 35, 36, 37, 38 and 39. This operation transforms 1s into 2s, 2s into 1s and maintains the 0 values, the matrix thereby having the same number (thirteen) of 0s, 1s and 2s in the first row:

| 1 | 2 | 3 | 4 | 5 | 6 | 7 | 8 | 9 | 10 | 11 | 12 | 13 | 14 | 15 | 16 | 17 | 18 | 19 | 20 | 21 | 22 | 23 | 24 | 25 | 26 | 27 | 28 | 29 | 30 | 31 | 32 | 33 | 34 | 35 | 36 | 37 | 38 | 39 |
|---|---|---|---|---|---|---|---|---|---|---|---|---|---|---|---|---|---|---|---|---|---|---|---|---|---|---|---|---|---|---|---|---|---|---|---|---|---|---|
| **0** | **0** | **0** | **0** | **0** | **0** | **0** | **0** | **0** | **0** | **0** | **0** | **0** | **1** | **1** | **1** | **1** | **1** | **1** | **1** | **1** | **1** | **1** | **1** | **1** | **1** | **2** | **2** | **2** | **2** | **2** | **2** | **2** | **2** | **2** | **2** | **2** | **2** | **2** |
| 0 | 0 | 0 | 0 | 1 | 1 | 1 | 1 | 1 | 1 | 1 | 1 | 1 | 0 | 0 | 0 | 0 | 0 | 0 | 0 | 0 | 0 | 1 | 1 | 1 | 1 | 2 | 2 | 2 | 2 | 2 | 1 | 1 | 1 | 1 | 1 | 1 | 1 | 1 |
| 0 | 1 | 1 | 1 | 0 | 0 | 0 | 1 | 1 | 1 | 2 | 2 | 2 | 0 | 0 | 0 | 1 | 1 | 1 | 2 | 2 | 2 | 0 | 0 | 0 | 1 | 2 | 2 | 1 | 1 | 1 | 0 | 0 | 0 | 2 | 2 | 2 | 1 | 1 |
| 1 | 0 | 1 | 2 | 0 | 1 | 2 | 0 | 1 | 2 | 0 | 1 | 2 | 0 | 1 | 2 | 0 | 1 | 2 | 0 | 1 | 2 | 0 | 1 | 2 | 0 | 2 | 1 | 0 | 2 | 1 | 0 | 2 | 1 | 0 | 2 | 1 | 0 | 2 |

The second row now contains thirteen 0s, twenty one 1s and five 2s. In order that it contain thirteen 1s and 2s, we will multiply by 2 module 3 the 8 vectors furthest to the right containing a 1 in the second row and an 0 in the previous row, i.e. the vectors or columns 6, 7, 8, 9, 10, 11, 12 and 13, the matrix thereby having the same number of 0s, 1s and 2s in the second row:

| 1 | 2 | 3 | 4 | 5 | 6 | 7 | 8 | 9 | 10 | 11 | 12 | 13 | 14 | 15 | 16 | 17 | 18 | 19 | 20 | 21 | 22 | 23 | 24 | 25 | 26 | 27 | 28 | 29 | 30 | 31 | 32 | 33 | 34 | 35 | 36 | 37 | 38 | 39 |
|---|---|---|---|---|---|---|---|---|---|---|---|---|---|---|---|---|---|---|---|---|---|---|---|---|---|---|---|---|---|---|---|---|---|---|---|---|---|---|
| **0** | **0** | **0** | **0** | **0** | **0** | **0** | **0** | **0** | **0** | **0** | **0** | **0** | **1** | **1** | **1** | **1** | **1** | **1** | **1** | **1** | **1** | **1** | **1** | **1** | **1** | **2** | **2** | **2** | **2** | **2** | **2** | **2** | **2** | **2** | **2** | **2** | **2** | **2** |
| **0** | **0** | **0** | **0** | **1** | **2** | **2** | **2** | **2** | **2** | **2** | **2** | **2** | **0** | **0** | **0** | **0** | **0** | **0** | **0** | **0** | **0** | **1** | **1** | **1** | **1** | **2** | **2** | **2** | **2** | **2** | **1** | **1** | **1** | **1** | **1** | **1** | **1** | **1** |
| 0 | 1 | 1 | 1 | 0 | 0 | 0 | 2 | 2 | 2 | 1 | 1 | 1 | 0 | 0 | 0 | 1 | 1 | 1 | 2 | 2 | 2 | 0 | 0 | 0 | 1 | 2 | 2 | 1 | 1 | 1 | 0 | 0 | 0 | 2 | 2 | 2 | 1 | 1 |
| 1 | 0 | 1 | 2 | 0 | 2 | 1 | 0 | 2 | 1 | 0 | 2 | 1 | 0 | 1 | 2 | 0 | 1 | 2 | 0 | 1 | 2 | 0 | 1 | 2 | 0 | 2 | 1 | 0 | 2 | 1 | 0 | 2 | 1 | 0 | 2 | 1 | 0 | 2 |

The third row now contains thirteen 0s, fifteen 1s and eleven 2s. In order that it contain the same number (thirteen) of 1s and 2s, we will multiply by 2 module 3 the two vectors furthest to the right that contain a 1 in the third row and a 0 in the previous rows, i.e., vectors or columns 3 and 4, the matrix thereby having the same number of 0s, 1s and 2s in all four rows:

| 1 | 2 | 3 | 4 | 5 | 6 | 7 | 8 | 9 | 10 | 11 | 12 | 13 | 14 | 15 | 16 | 17 | 18 | 19 | 20 | 21 | 22 | 23 | 24 | 25 | 26 | 27 | 28 | 29 | 30 | 31 | 32 | 33 | 34 | 35 | 36 | 37 | 38 | 39 |
|---|---|---|---|---|---|---|---|---|---|---|---|---|---|---|---|---|---|---|---|---|---|---|---|---|---|---|---|---|---|---|---|---|---|---|---|---|---|---|
| **0** | **0** | **0** | **0** | **0** | **0** | **0** | **0** | **0** | **0** | **0** | **0** | **0** | **1** | **1** | **1** | **1** | **1** | **1** | **1** | **1** | **1** | **1** | **1** | **1** | **1** | **2** | **2** | **2** | **2** | **2** | **2** | **2** | **2** | **2** | **2** | **2** | **2** | **2** |
| **0** | **0** | **0** | **0** | **1** | **2** | **2** | **2** | **2** | **2** | **2** | **2** | **2** | **0** | **0** | **0** | **0** | **0** | **0** | **0** | **0** | **0** | **1** | **1** | **1** | **1** | **2** | **2** | **2** | **2** | **2** | **1** | **1** | **1** | **1** | **1** | **1** | **1** | **1** |
| **0** | **1** | **2** | **2** | **0** | **0** | **0** | **2** | **2** | **2** | **1** | **1** | **1** | **0** | **0** | **0** | **1** | **1** | **1** | **2** | **2** | **2** | **0** | **0** | **0** | **1** | **2** | **2** | **1** | **1** | **1** | **0** | **0** | **0** | **2** | **2** | **2** | **1** | **1** |
| **1** | **0** | **2** | **1** | **0** | **2** | **1** | **0** | **2** | **1** | **0** | **2** | **1** | **0** | **1** | **2** | **0** | **1** | **2** | **0** | **1** | **2** | **0** | **1** | **2** | **0** | **2** | **1** | **0** | **2** | **1** | **0** | **2** | **1** | **0** | **2** | **1** | **0** | **2** |

In addition to indicating which coins must appear in each weighing, this matrix identifies with the value 1 those coins which should appear, for example, on the left side of the scales and with a 2 those which must appear on the right and with 0 those which do not intervene in the weighing. The following weighings can therefore be deduced:

**$s_1$ 14, 15, 16, 17, 18, 19, 20, 21, 22, 23, 24, 25, 26 <> 27, 28, 29, 30, 31, 32, 33, 34, 35, 36, 37, 38, 39**
**$s_2$ 5, 23, 24, 25, 26, 32, 33, 34, 35, 36, 37, 38, 39 <> 6, 7, 8, 9, 10, 11, 12, 13, 27, 28, 29, 30, 31**
**$s_3$ 2, 11, 12, 13, 17, 18, 19, 26, 29, 30, 31, 38, 39 <> 3, 4, 8, 9, 10, 20, 21, 22, 27, 28, 35, 36, 37**
**$s_4$ 1, 4, 7, 10, 13, 15, 18, 21, 24, 28, 31, 34, 37 <> 3, 6, 9, 12, 16, 19, 22, 25, 27, 30, 33, 36, 39**

The result of the weighings $s_1$, $s_2$, $s_3$, $s_4$ obtained by assigning 1 to the left-hand tilt, 0 for balance and 2 for the right-hand tilt will serve to identify in the matrix the coin that has in its column the values $s_1$, $s_2$, $s_3$, $s_4$ and in this case that it is heavier.

If $s_1$, $s_2$, $s_3$, $s_4$ cannot be found in the matrix, the vector $s_1/2$, $s_2/2$, $s_3/2$, $s_4/2$, will be obtained (this operation is equivalent to replacing value 1 with 2 and 2 with 1 and 0 remaining unchanged) the coin with this value in its column will be identified in the matrix and in this case it is lighter.

Example 1
Supposing the result of the four weighings is:

$s_1=1$, $s_2=0$, $s_3=0$, $s_4=0$

The counterfeit coin is 14 and it is heavier

Example 2:

Supposing the result of the four weighings is:

$s_1=1$, $s_2=1$, $s_3=0$, $s_4=2$

The counterfeit coin is 25 and it is heavier.

Example 3:

Supposing the result of the four weighings is:

$s_1=2$, $s_2=2$, $s_3=0$, $s_4=1$

This column is not in the matrix and it is therefore lighter and

$s_1=2/2=1$, $s_2=2/2=1$, $s_3=0/2=0$, $s_4=1/2=2$

Corresponds to coin 25.

## *11. Extension of the procedure to identify more than one counterfeit coin.*

When there is more than one counterfeit coin, it will be necessary to choose a base 3 code capable of detecting and correcting as many errors as coins. The check matrix of this code shows both how to perform the weighings and how to identify coins by their weight excess or deficiency, as previously.

## *12. Bibliography*

(1) C. Shannon"The Mathematical Theory of Communication" *Bell System Tech*. J.,**27**,(1948)

(2) J. Domínguez-Montes,"Randomness and Particle Size", *Physics Essays* Vol **18**,1,81-94 (March 2005)..

(3) A.M. Yaglon and I.M. Yaglon, "*Probabilité et Information*", Dunod (1969).

(4) J. Domínguez-Montes,"Solución definitiva al problema de la moneda falsa" *Qüestiió* **7,**2,451-458, (Jun 1983)

(5) F.M. Reza, "*An introduction to information theory*". McGRAW-HILL (1961).

(6) W.W. Peterson –E.J. Weldon, "*Error-correcting codes*". THE MIT PRESS (1961).